\documentclass{JINST}
\usepackage{graphicx}
\usepackage{epstopdf}
\usepackage{subfigure}
\usepackage{capt-of}

\title{The Impact of Photon Flight Path on S1 Pulse Shape Analysis in Liquid Xenon Two-phase Detectors}
\author{M. Moongweluwan$^a$ for the LUX collaboration\\
\llap{$^a$}Department of Physics and Astronomy, University of Rochester,\\
Rochester, New York 14627,  USA\\
E-mail: \email{mmoongwe@pas.rochester.edu}}
\abstract{
	The LUX dark matter search experiment is a 350 kg dual-phase xenon time projection chamber located at the 4850 ft level of the Sanford Underground Research Facility in Lead, SD. The success of two-phase xenon detectors for dark matter searches relies on their ability to distinguish electron recoil (ER) background events from nuclear recoil (NR) signal events. Typically, the NR-ER discrimination is obtained from the ratio of the electroluminescence light (S2) to the prompt scintillation light (S1). Analysis of the S1 pulse shape is an additional discrimination technique that can be used to distinguish NR from ER. Pulse-shape NR-ER discrimination can be achieved based on the ratio of the de-excitation processes from singlet and triplet states that generate the S1. The NR S1 is dominated by the de-excitation process from singlet states with a time constant of about 3 ns while the ER S1 is dominated by the de-excitation process from triplet states with a time constant of about 24 ns. As the size of the detectors increases, the variation in the S1 photon flight path can become comparable to these decay constants, reducing the utility of pulse-shape analysis to separate NR from ER. The effect of path length variations in the LUX detector has been studied using the results of simulations and the impact on the S1 pulse shape analysis is discussed.
}


\begin{document}
	\section{Introduction} \label{LUXIntro}
	The Large Underground Liquid Xenon Experiment (LUX) is a WIMP dark matter search experiment using a dual-phase xenon detector, described in detail in \mbox{Ref. \cite{LUXNimPaper}.} The basic principle of operation is that an energy deposition from an incident particle produces prompt scintillation light, called S1, and ionization electrons. The electrons drift to the liquid surface due to an applied electric field and are extracted into the gas region to produce electroluminescence light, called S2. The S1 and S2 light is observed with 122 photomultiplier tubes (PMTs) separated into two arrays of 61 PMTs each, located at the top and bottom of the detector. The LUX cryostat consists of two concentric titanium cans. The xenon is placed inside the inner can; a vacuum space is created between the outer and the inner cans. The inner wall of the LUX time-projection chamber (TPC) is made with polytetrafluoroethylene (PTFE) to enhance light collection efficiency due to its high reflectivity at xenon scintillation wavelengths. The TPC is a dodecagon in shape, about 60 cm in height from top to bottom PMT arrays, and about 49 cm in diameter. The liquid xenon surface is about 54.4 cm above the bottom PMT array. The gaseous phase spans a region of about 5 cm between the liquid surface and the top PMT array. LUX has been operating at the Sanford Underground Research Facility (SURF) in Lead, SD, since Oct 2012. The operation of liquid xenon detectors for dark matter searches relies on their ability to distinguish electron recoil (ER) background events from nuclear recoil (NR) signal events \cite{LUXRun3WS, ReviewAprille, ReviewAraujo}. Typically, the NR-ER discrimination is obtained from the ratio of S2 light to S1 light. NR events produce less S2 light than ER for S1 signals of the same size. In addition to the S2 to S1 ratio, NR-ER discrimination can also be obtained based on the S1 photon timing distribution \cite{PSD_XMass, PSD_Kwong}. The energy deposition on the xenon atom from the recoil produces an excited positive atom and ionization electrons. Two physics processes contribute to the S1 photon timing. The first process is a diatomic de-excitation process. Excited atoms from the recoil form an excited diatomic xenon molecule with a nearby xenon atom, then undergoes a de-excitation process to release photons. The de-excitation can occur from a singlet state, a state which has a short lifetime, or from a triplet state, which has a longer lifetime. The other process is a recombination between the ionization electrons and the positive atom. This process produces an excited xenon atom which eventually forms an excited diatomic xenon molecule and undergoes the de-excitation process. The ratio of excitation to ionization, as well as the ratio of singlet excitation to triplet excitation, is different for NR and ER events. Therefore, the timing distribution of the S1 photons can be used to discriminate NR from ER \cite{ReviewAprille, ReviewAraujo}. 

	\section{Simulation method}
	
\begin{figure}
		\begin{minipage}{0.58\linewidth}
			\begin{tabular}{|l|l|}
				\hline Quantity & Values \\ 
				\hline Singlet lifetime & 3.1 \(\pm\) 0.7 ns \\ 
				\hline Triplet lifetime & 24 \(\pm\) 1 ns \\ 
				\hline Singlet to triplet probability ratio - NR & 1.6 \(\pm\) 0.2 \\ 
				\hline Singlet to triplet probability ratio - ER & 0.6 \(\pm\) 0.2 \\ 
				\hline 
			\end{tabular}
			\captionof{table}{S1 timing distribution parameters. The NR and ER singlet to triplet ratios are values at 0 kV/cm and 4 kV/cm electric field, respectively \cite{S1Nest, Hitachi1983, Kubota1978}.}%
			\label{table_1}
		\end{minipage}
		\hfill
		\begin{minipage}{0.40\linewidth}
			\centering
			\includegraphics[width=0.8\linewidth]{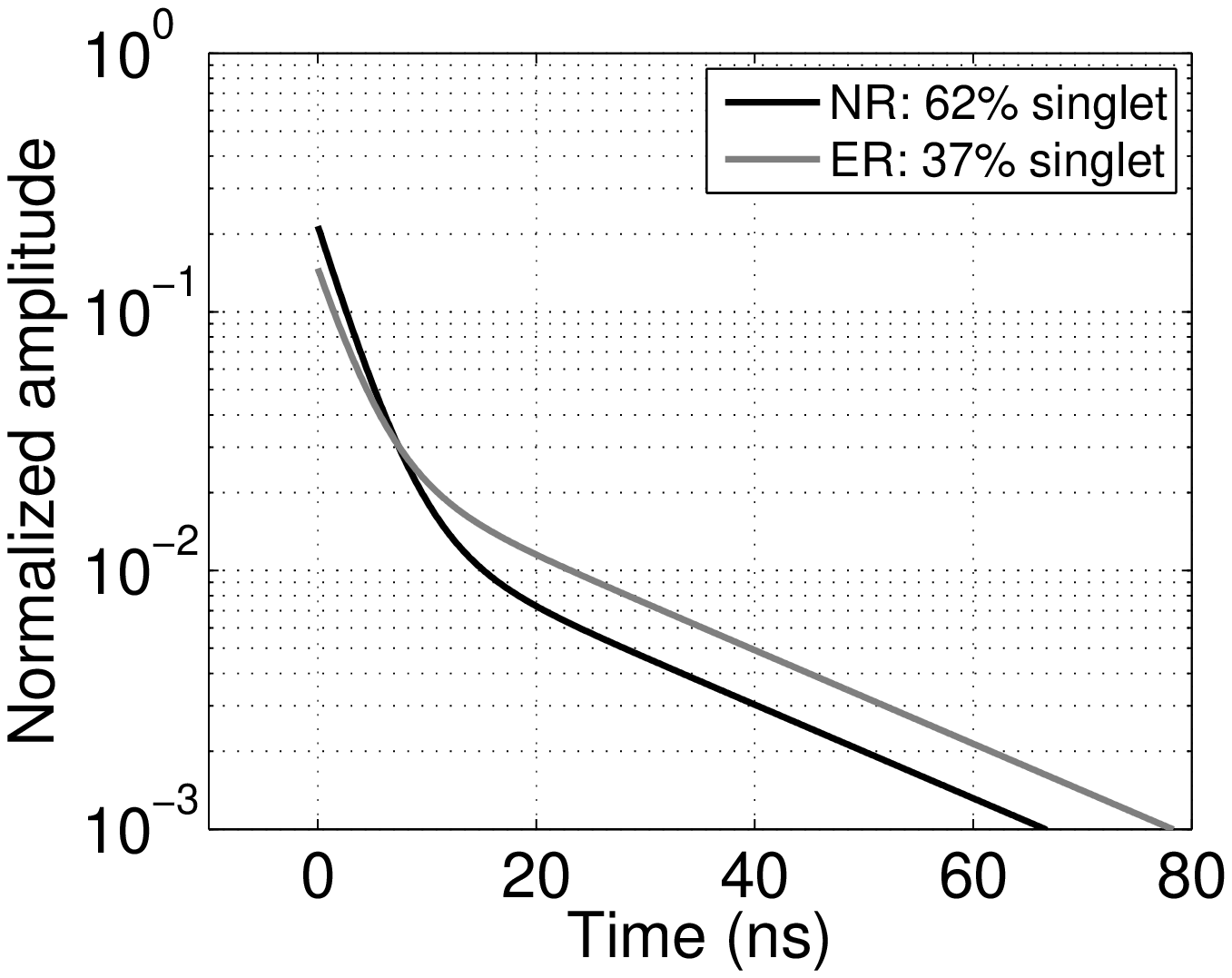}
			\captionof{figure}{S1 photon timing distribution of NR S1 (black) and ER S1 (gray) according to the parameters in Table \protect\ref{table_1}}
			\label{fig:IdealS1}
		\end{minipage}
	\end{figure}

	To study the S1 pulse shape, a simulation is carried out using the LUXSim Monte Carlo simulation package, based on Geant4.9.4.p04 \cite{Geant4_1, Geant4_2, LUXSim}. We use the S1 model as described in \cite{S1Nest, Hitachi1983, Kubota1978}. The parameters responsible for the S1 photon timing distribution from measurements are quoted in Table \ref{table_1} and the plot of S1 photon timing distributions according to these parameters is shown in Fig. \ref{fig:IdealS1}. We use the singlet to triplet ratio of ER that corresponds to the case when the recombination process is not completely suppressed. The time constant is a function of recoil energy and applied electric field. In LUX, the time constant is estimated to be in the order of a several nanoseconds for the energy range of interest \cite{S1Nest}. However, the explicit recombination time constant in ER is not included in the study at the moment. In NR, the time constant is negligible \cite{S1Nest}. The uncertainties associated to each parameter are handled as statistical fluctuations. The fluctuation of each parameter is assumed to be a Gaussian distribution with a standard deviation of the respective uncertainty of that parameter. Both NR and ER singlet to triplet ratios are assumed to be independent of the recoil energy. The xenon scintillation wavelength is assumed to be 177 nm \cite{XeWavelength}. 
	
\begin{figure}
		\centering
		\includegraphics[width=0.48\linewidth]{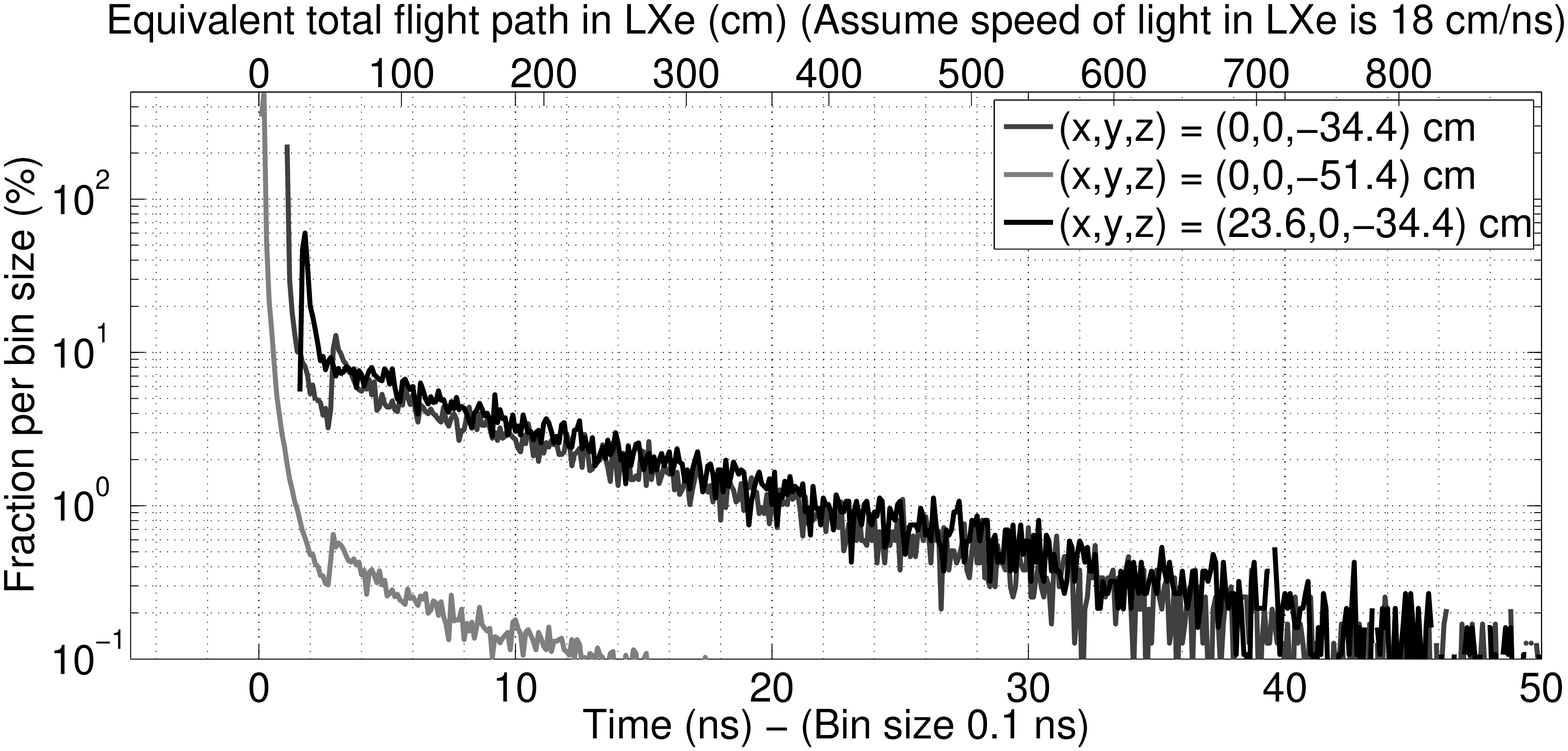}
		\caption{Photon timing distribution from the center bottom PMT for three different photon source locations. \((x,y,z) = (0,0,0)\) cm is the center of the detector at the liquid surface. The shortest time corresponds to photons traveling in a straight path from the source to the PMT. Some of the photons can take a longer path by scattering with a xenon atom via Rayleigh scattering and produce the exponential tail. The sharp rise between 2 and 3 ns is associated with photons that reflect once from the TPC wall.}
		\label{fig:FlightPathEffect}
	\end{figure}

	To compare the time distributions shown in Fig. \ref{fig:IdealS1} to the measured distributions, detector effects must be taken into account. Due to the size of the detector, a photon can take many different paths between the point of origin and a PMT. To demonstrate this flight path variation effect, a simulation is done by emitting 177 nm photons isotropically from a specific source location. The photons are propagated through the detector until they reach the PMTs. The time each photon takes to reach a PMT is recorded. This process is repeated for several source locations uniformly distributed within a cylindrical region of 20 cm radius, and a height between 2.4 and 48.5 cm with respect to the bottom PMT array. A selected result is shown in Fig. \ref{fig:FlightPathEffect}. These simulations show that many photons take more than 3.5 ns, the time of flight between the top and bottom PMT arrays, to arrive at any given PMT. This indicates that many photons travel in a non-straight path from the point of origin to the PMT. 
	
	Furthermore, each photoelectron generated at the PMT photocathode can take a different time to travel to the anode of the PMT. This transit time spread is a function of PMT bias voltage. LUX uses Hamamatsu R8778 PMTs which have a mean transit time of 41 ns and the transit time spread of 4 ns (FWHM) at 1500 V \cite{PMTSpec, LUXPMTPaper}. The PMT mean transit time variation due to different bias voltages is not included in the simulation, since it can be corrected for. 
	
\begin{figure}
		\centering
		\subfigure{\includegraphics[width=0.48\linewidth]{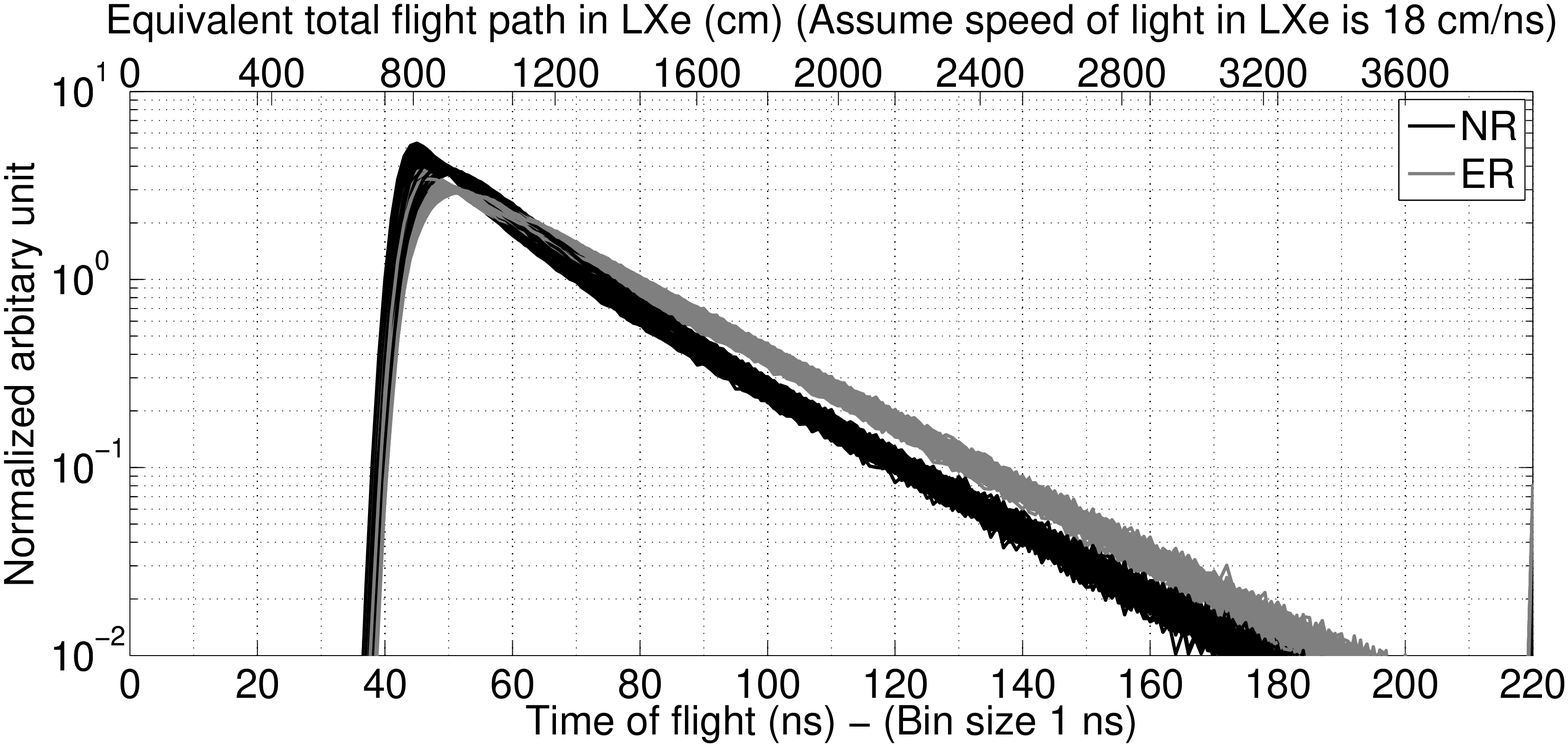}}
		\hfill
		\subfigure{\includegraphics[width=0.48\linewidth]{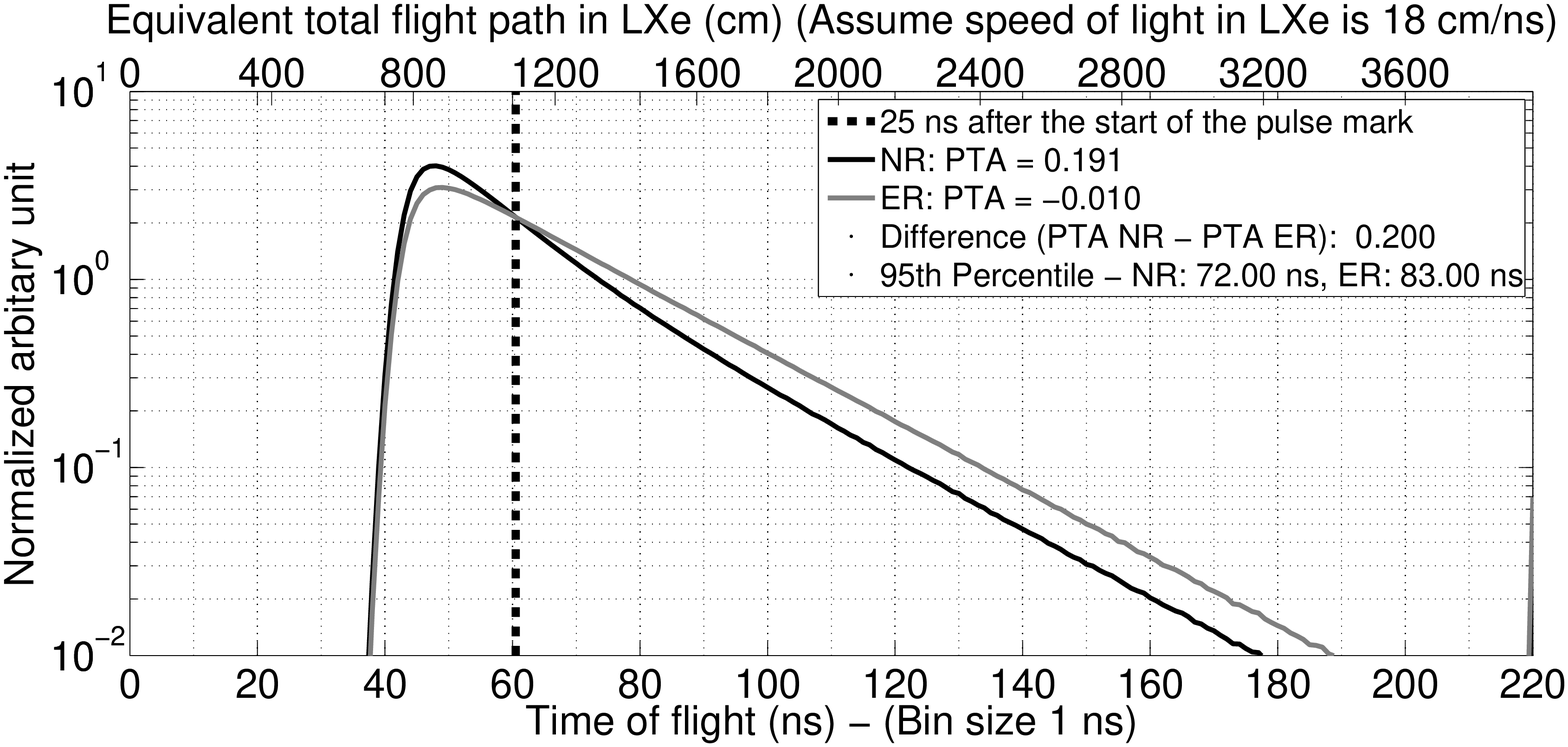}}
		\caption{Left: S1 photon timing distribution including effects from flight path variation and PMT transit time spread. Both NR and ER consist of several curves representing S1 distributions from different locations. Right: Average S1 photons timing distribution from all locations shown in the left plot.}
		\label{fig:S1TimingDist}
	\end{figure}

	The S1 pulse shape generated from different locations including the effect from flight path variation and PMT transit time spread is shown in Fig. \ref{fig:S1TimingDist}. 
		
	\section{S1 pulse shape discrimination}
	The results show that the observed S1 photon timing distributions \mbox{(Fig. \ref{fig:S1TimingDist})} are different from the distribution of the S1 photon generation model alone (Fig. \ref{fig:IdealS1}). These results show that the spread in the photon timing from the flight path variation and the PMT transit time spread will degrade NR-ER discrimination. The impact is location-dependent and will be greater as the size of the detector increases. Nevertheless, the results suggest that NR-ER discrimination can be achieved, even without position-dependent corrections. These corrections will improve the discrimination. The averaged S1 photon timing distribution, obtained from averaging the distributions from all locations, is shown in Fig. \ref{fig:S1TimingDist} (right). To quantify the discrimination, we use Photon Timing Asymmetry (PTA) parameter, defined in Eq. (\ref{eq:1});
	\begin{equation}	
	\label{eq:1}
	PTA = \frac{(Number\:of\:photons\:between\:0\:and\:25\:ns)\:-\:(Number\:of\:photons\:between\:25\:and\:200\:ns)}{(Number\:of\:photons\:between\:0\:and\:25\:ns)\:+\:(Number\:of\:photons\:between\:25\:and\:200\:ns)}
	\end{equation}	
	In Eq. (\ref{eq:1}), $t\:=\:0$ is defined as the time of the first observed photon. The number of 25 ns, which is the number where the curves in Fig. \ref{fig:S1TimingDist} (right) cross, is chosen so that the difference between the PTA of NR and ER is maximized.
		
\begin{figure}
		\begin{minipage}[t]{.48\textwidth}
			\centering
			\includegraphics[width=0.9\linewidth]{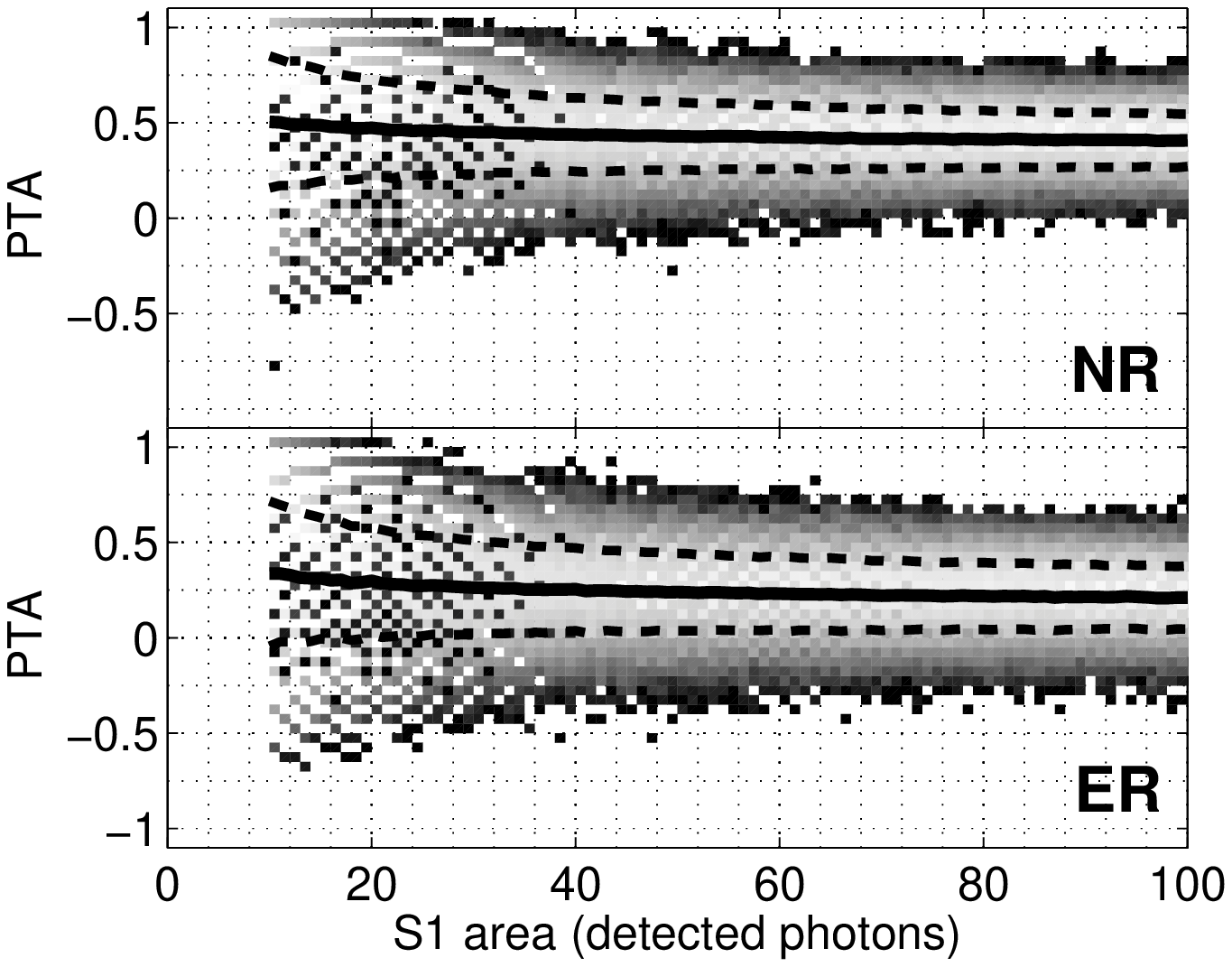}
			\caption{PTA bands from simulated NR S1 (top) and ER S1 (bottom). Both are overlaid with their respective mean (solid line) and \(\pm\)1.28\(\sigma\) (dashed line). The discrete pattern at low S1 area is due to limited possible outcomes from the definition of PTA.}
			\label{fig:DiscBand3}
		\end{minipage}%
		\hfill
		\begin{minipage}[t]{.48\textwidth}
			\centering
			\includegraphics[width=0.9\linewidth]{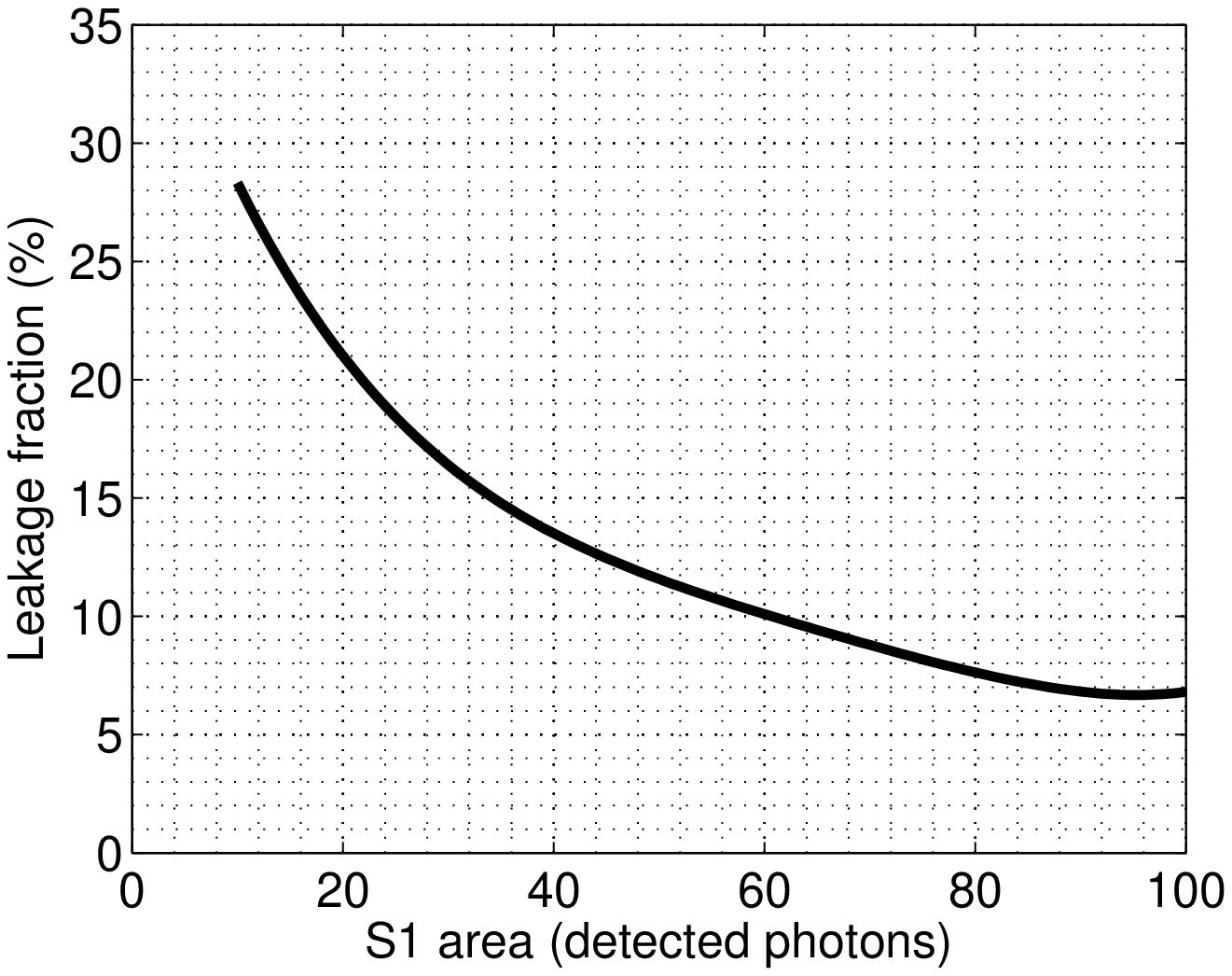}
			\caption{Leakage fraction calculated from acceptance region defined.}
			\label{fig:Leakage}
		\end{minipage}
	\end{figure}

	The NR-ER discrimination is studied for S1s with 10 to 100 detected photons. S1s of different sizes are generated from several locations within the same cylindrical region. The PTA is calculated for each S1 individually and plotted as a function of S1 size. For each S1 bin, the mean and standard deviation of the PTA are calculated. The results are shown in Fig. \ref{fig:DiscBand3}. The PTA bands from NR and ER are clearly distinguishable. We define the NR acceptance region as the region above the NR band mean. The ER leakage fraction into the NR acceptance region is calculated and shown in \mbox{Fig. \ref{fig:Leakage}}. This acceptance region yields a discrimination power of \textasciitilde72-93\% over an S1 range of 10-100 detected photons. The method can be applied concurrently with the typical ratio of the S2 light over the S1 light for better overall discrimination.

	\section{Conclusions}
	In this paper, we have discussed detector effects, especially the photon flight path variation and the PMT transit time spread, on the S1 photon timing distribution. These effects cause the observed S1 photon timing distribution to be different from the actual S1 photon timing distribution. We have introduced a discrimination parameter, the PTA, and studied the NR-ER discrimination using PTA based on the S1 model and the LUX detector geometry. The discrimination power is \textasciitilde72-93\% for S1 sizes of 10-100 detected photons. This pulse shape analysis can be applied to the data concurrently with the S2 to S1 ratio to improve the overall discrimination.
	\acknowledgments
	This work was partially supported by the U.S. Department of Energy (DOE) under award numbers DE-FG02-08ER41549, DE-FG02-91ER40688, DE-FG02-95ER40917, DE-FG02-91ER40674, DE-NA0000979, DE-FG02-11ER41738, DE-SC0006605, DE-AC02-05CH11231, DE-AC52-07NA27344, and DE-FG01-91ER40618; the U.S. National Science Foundation under award numbers PHYS-0750671, PHY-0801536, PHY-1004661, PHY-1102470, PHY-1003660, PHY-1312561, PHY-1347449; the Research Corporation grant RA0350; the Center for Ultra-low Background Experiments in the Dakotas (CUBED); and the South Dakota School of Mines and Technology (SDSMT). LIP-Coimbra acknowledges funding from Fundação para a Ciência e Tecnologia (FCT) through the project-grant CERN/FP/123610/2011. Imperial College and Brown University thank the UK Royal Society for travel funds under the International Exchange Scheme (IE120804). The UK groups acknowledge institutional support from Imperial College London, University College London and Edinburgh University, and from the Science \& Technology Facilities Council for PhD studentship ST/K502042/1 (AB). The University of Edinburgh is a charitable body, registered in Scotland, with registration number SC005336.

	\clearpage
	

\begin{thebibliography}{1}
		\bibitem{LUXNimPaper} LUX Collaboration, D.\,S.\,Akerib, et al., {\it The large underground xenon (LUX) experiment}, {\it Nucl. Instr. Meth.}  {\bf A 704} (2013) 111.
		\bibitem{LUXRun3WS} LUX Collaboration, D.\,S.\,Akerib, et al., {\it First results from the LUX dark matter experiment at the Sanford underground research facility}, {\it Phys. Rev. Lett.} {\bf 122} (2013) 091303.
		\bibitem{PSD_XMass} XMASS Collaboration, K.\,Ueshima, et al., {\it Scintillation-only based pulse shape discrimination for nuclear and electron recoils in liquid xenon}, {\it Nucl. Instr. Meth.}  {\bf A 659} (2011) 161.
		\bibitem{PSD_Kwong} J.\,Kwong, et al., {\it Scintillation pulse shape discrimination in a two-phase xenon time projection chamber}, {\it Nucl. Instr. Meth.} {\bf A 612} (2010) 328.
		\bibitem{ReviewAprille} A.\,Aprile, T.\,Doke, {\it Liquid xenon detectors for particle physics and astrophysics}, {\it Rev. Mod. Phys.} {\bf 82} (2015) 2053 
		\bibitem{ReviewAraujo} V.\,Chepel, H.\,Araujo, {\it Liquid noble gas detectors for low energy particle physics}, 2013 {\it JINST} {\bf 8}  R04001.
		\bibitem{Geant4_1} Geant4 collaboration, S.\,Agostinelli, et al., {\it Geant4: A simulation toolkit}, {\it Nucl. Instrum. Meth. } {\bf A 506} (2003) 250
		\bibitem{Geant4_2} Geant4 collaboration, J.\,Allison, et al., {\it Geant4 developments and applications}, {\it Nuclear Science, IEEE Transactions on } {\bf 53} (2005) 270 
		\bibitem{LUXSim} LUX Collaboration, D.\,S.\,Akerib, et al., {\it LUXSim: A component-centric approach to low-background simulations}, {\it Nucl. Instr. Meth.} {\bf A 675} (2011) 63.
		\bibitem{S1Nest} J.\,Mock, et al., {\it Modeling pulse characteristics in xenon with NEST} 2014 {\it JINST} {\bf 9} T04002. 
		\bibitem{Hitachi1983} A.\,Hitachi, et al., {\it Effect of ionization density on the time dependence of luminescence from liquid argon and xenon}, {\it Phys. Rev.} {\bf B 27}, 5279 (1983)
		\bibitem{Kubota1978} S.\,Kubota, M.\,Hishida and J.\,Z.\,Ruan, {\it Evidence for a triplet state of the self-trapped exciton states in liquid argon, krypton and xenon}, {\it J. Phys.} {\bf C 11} (1978) 2645.
		\bibitem{XeWavelength} J.\,Jortner, et al., {\it Localized excitations in condensed Ne, Ar, Kr, and Xe}, {\it J. Chem Phys.} {\bf 42} (1965) 4250 
		\bibitem{PMTSpec} Hamamatsu Photonics K.K.,	{\it R8778 data sheet}, December 2011.
		\bibitem{LUXPMTPaper} LUX Collaboration, D.\,S.\,Akerib, et al., {\it An ultra-low background PMT for liquid xenon detectors}, {\it Nucl. Instr. Meth.} {\bf A 703} (2012) 1.		
	\end{thebibliography}
\end{document}